\documentclass[11pt]{article}
\usepackage[left=1in,top=1in,right=1in,bottom=1in,head=.1in,nofoot]{geometry}

\usepackage{setspace,url,bm,amsmath} 
\usepackage{float}
\usepackage{caption}
\usepackage{hyperref}
\usepackage{algorithm}
\usepackage{algpseudocode}
\usepackage{enumitem}
\usepackage{subcaption,siunitx,booktabs}

\usepackage{titlesec} 
\usepackage{rotating}
\titlelabel{\thetitle.\quad} 
\titleformat*{\section}{\bf\large}

\usepackage{graphicx} 
\usepackage{bbm}
\usepackage{latexsym}
\usepackage{color}
\usepackage{authblk}
\usepackage{amsthm}
\usepackage{amsfonts}
\usepackage{bm}
\usepackage{multirow}
\usepackage{xcolor}
\usepackage{comment}

\newtheorem{theorem}{Theorem}

\newtheorem{lemma}{Lemma}
\newtheorem{corollary}{Corollary}

\theoremstyle{definition}
\newtheorem{definition}{Definition}
\newtheorem{example}{Example}
\newtheorem{remark}{Remark}

\usepackage{natbib} 

\newcommand{\T}{\text{T}}
\newcommand{\trueparm}{\bm \theta}
\newcommand{\trueparmi}{\gamma_i}
\newcommand{\anytrueparm}{\Tilde{\bm \theta}}

\newcommand{\parmspace}{\mathbf{\Theta}}
\newcommand{\outp}{y}
\newcommand{\outpi}{y_i}
\newcommand{\vecyj}[1]{\bm y_{#1}}
\newcommand{\inp}{\mathbf{X}}
\newcommand{\inppart}[1]{\mathbf{X_{#1}}}
\newcommand{\inpj}[1]{\mathbf{x_#1}}
\newcommand{\inpjt}[1]{\mathbf{x^{\T}_#1}}

\newcommand{\err}{\bm{\epsilon}}
\newcommand{\erri}{\epsilon_i}
\newcommand{\sigmasq}{\sigma^2}
\newcommand{\inps}{\mathbf{x}}

\newcommand{\likelihood}{f( y | \trueparm, \mathbf{x})}
\newcommand{\likelihoodi}[1]{f(y_i | \trueparm, \phi, \inpj{#1})}
\newcommand{\growthfn}[1]{g ( \trueparm, #1 )}
\newcommand{\gderiv}[1]{\bm{g^\prime ( \trueparm, #1 )}}
\newcommand{\const}{\zeta}
\newcommand{\linfn}{{\inp}^{\T} \trueparm}
\newcommand{\istarinv}{\mathbf{I}^{*-1}}
\newcommand{\istar}{\mathbf{I}^{*}}

\newcommand{\totalfisher}[3]{\sum_{#2=1}^#1 \mathbf{I}\left(#3, \inpj{#2}\right)}
\newcommand{\loglik}{\log  \likelihood }
\newcommand{\anyloglik}{\log  f(\bm{y} | \Tilde{\trueparm}, \inp) }

\newcommand{\score}{\nabla \log \likelihood}
\newcommand{\Rcal}{\mathcal{R}}
\newcommand{\exptspace}{\bm{\Omega}}
\newcommand{\designspace}{\bm{\Xi}}
\newcommand{\anydesign}{\mathbf{\xi}}
\newcommand{\initialdesign}{\mathbf{\xi}_0}
\newcommand{\weights}{w_1, w_2, \ldots, w_k}
\newcommand{\optdesign}{\mathbf{\xi^*}}
\newcommand{\initialrun}{n_1}
\newcommand{\points}[2]{#1_1, #1_2, \ldots, #1_{#2}}
\newcommand{\pointsbf}[2]{\mathbf{#1}_1, \mathbf{#1}_2, \ldots, \mathbf{#1}_{#2}}
\newcommand{\designsize}{n}
\newcommand{\blncdesignsize}{k}
\newcommand{\numsim}{r}

\newcommand{\mle}[1]{\widehat{\trueparm}_{#1}}
\newcommand{\optimfn}[3]{\mathbf{\Psi}(#3{#1} , \inp_{#2})}

\newcommand{\optptsmain}[2]{\mathbf{x^*_#2(#1)}}
\newcommand{\optpts}[2]{x^*_#2(#1)}
\newcommand{\opttrue}[1]{x^*(#1)}
\newcommand{\liparg}[3]{\lvert \optpts{#1}{#2} - \opttrue{#3} \rvert}
\newcommand{\liprhs}[2]{\lvert #1 - #2 \rvert}
\newcommand{\gexpo}{\alphaone \exp^{-\alphatwo /x}}
\newcommand{\glin}{\intercept + \slope x}
\newcommand{\Xmin}{x_{\text{min}}}
\newcommand{\Xmax}{x_{\text{max}}}
\newcommand{\changept}{x_0}
\newcommand{\lipsch}{L}
\newcommand{\alphaone}{\alpha_1}
\newcommand{\alphatwo}{\alpha_2}
\newcommand{\intercept}{a}
\newcommand{\slope}{b}
\newcommand{\modeltrueparmval}[2]{{#1}^*_{#2}}

\newcommand{\scorerhs}{G (\outp , \trueparm , \mathbf{x})}
\newcommand{\hess}{H (\outp , \trueparm , \mathbf{x})}
\newcommand{\nbd}{N(\trueparm)}
\newcommand{\K}{K (\outp, \trueparm, \mathbf{x})}
\newcommand{\dimn}{d}
\newcommand{\sigmafield}[2]{\mathcal{F}_{#1, #2}}
\newcommand{\gradient}[2]{\sum_{#1 = 1}^{#2} G (Y_{#1} , \trueparm , \mathbf{X}_{#1})}
\newcommand{\bigO}[2]{O(#1^{#2})}
\newcommand{\hessplusfish}[2]{\sum_{#1 = 1}^{#2} \left \{ H (Y_{#1} , \trueparm , \mathbf{X}_{#1}) + \mathbf{I}\left(\trueparm, \mathbf{X}_#1\right) \right \} }
\newcommand{\hessian}[2]{\sum_{#1 = 1}^{#2} H (Y_{#1} , \trueparm , \mathbf{X}_{#1})}

\newcommand{\glmni}[2]{n_{#1#2}}
\newcommand{\glmpar}[1]{\beta_#1}
\newcommand{\piprob}{\pi}
\newcommand{\xmatrix}[1]{\mathcal{X}_{#1}}
\newcommand{\weightmat}[1]{\bm{W_{#1}}}
\newcommand{\pvec}{\bm{p}}
\newcommand{\vvec}{\bm{v}}
\newcommand{\wcomp}[2]{w_{#1#2}}
\newcommand{\woptcomp}[2]{w_{#1#2}^*}
\newcommand{\vcomp}[2]{v_{#1#2}}
\newcommand{\pcomp}[2]{p_{#1#2}}
\newcommand{\poptcomp}[2]{p_{#1#2}^*}
\newcommand{\poptvec}{\bm{p}^*}
\newcommand{\poptfn}[3]{p_{#1#2}^* (#3)}
\newcommand{\batchsize}{B}

\newcommand{\frobnorm}[1]{\left \|#1 \right \|_{\text{F}}}
\newcommand{\lipschtwo}[3]{\sum_{#2=1}^#1 \frobnorm{ \left \{\frac{1}{#1} \sum_{#2=1}^#1 \mathbf{I}\left(#3, \hat{X}_#2\right) - \frac{1}{#1} \sum_{#2=1}^#1 \mathbf{I}\left(#3, x^*_#2(#3)\right)  \right \}}}
\newcommand{\lipschtwonext}[3]{\sum_{#2=1}^#1 \frobnorm{ \left \{\mathbf{I}\left(#3, \hat{X}_#2\right) -  \mathbf{I}\left(#3, x^*_#2(#3)\right)  \right \}}}
\newcommand{\lipschinfo}{L_1}
\newcommand{\totalfisheropt}[3]{\sum_{#2=1}^#1 \mathbf{I}\left(#3, x^*_#2(#3)\right)}

\def\convergep{\stackrel{p}{\longrightarrow}}
\def\converged{\stackrel{d}{\longrightarrow}}
\def\drawiid{\stackrel{i.i.d.}{\sim}}

\newcommand{\gradientan}[2]{\sum_{#1=1}^{#2} \mathbb{E} \Big[G(Y_{#1}, \trueparm, \mathbf{X}_{#1})^2 \mathbb{I}_{(\left \lvert G(Y_{#1}, \trueparm, \mathbf{X}_{#1}) \right \rvert > \epsilon s_{#2})} \Big| \sigmafield{#2}{#1-1}\Big]}
\newcommand{\gradientantwo}[2]{\sum_{#1=1}^{#2} \mathbb{E} \left[G(Y_{#1}, \trueparm, \mathbf{X}_{#1})^2 | \sigmafield{#2}{#1-1}\right]}
\newcommand{\gradientest}[2]{\sum_{#1 = 1}^{#2} G (Y_{#1} , \mle{#2} , \mathbf{X}_{#1})}
\newcommand{\thirdpartial}[2]{\sum_{#1 = 1}^{#2} R (Y_{#1} , \trueparm' , \mathbf{X}_{#1})}

\newcommand{\wtltwo}[2]{\|#1\|_{#2}}
\newcommand{\estdesign}{\xi_n}

\newpage
\newcommand{\ct}{\textcolor{red}{XXXX}}
\newcommand{\thetahat}{\widehat{\bm \theta}}
\newcommand{\inProb}{\stackrel{p}{\rightarrow}}
\newcommand{\dist}{\mathbf{\xi}}
\newcommand{\Fish}{\mathbf{I}}
\newcommand{\opnorm}[1]{\| #1 \|_{\mathrm{op}}}
\newcommand{\Xdom}{\mathcal{X}}
\newcommand{\ParDom}{\Theta}
\newcommand{\Exs}{\mathbb{E}}
\newcommand{\Lip}{L}
\newpage 
\newcommand{\EFish}{\mathbf{J}}
\newcommand{\Var}{\mathrm{Var}}

\title{PICS: A sequential approach to obtain optimal designs for non-linear models leveraging closed-form solutions for faster convergence}
\author{Suvrojit Ghosh, Koulik Khamaru, Tirthankar Dasgupta}
\affil{Department of Statistics, Rutgers University}
\date{}

\begin{document}

\maketitle

\begin{abstract}
D-Optimal designs for estimating parameters of response models are derived by maximizing the determinant of the Fisher information matrix. For non-linear models, the Fisher information matrix depends on the unknown parameter vector of interest, leading to a weird situation that in order to obtain the D-optimal design, one needs to have knowledge of the parameter to be estimated. One solution to this problem is to choose the design points sequentially, optimizing the D-optimality criterion using parameter estimates based on available data, followed by updating the parameter estimates using maximum likelihood estimation. On the other hand, there are many non-linear models for which closed-form results for D-optimal designs are available, but because such solutions involve the parameters to be estimated, they can only be used by substituting “guestimates” of parameters. In this paper, a hybrid sequential strategy called PICS (Plug into closed-form solution) is proposed that replaces the optimization of the objective function at every single step by a draw from the probability distribution induced by the known optimal design by plugging in the current estimates. Under regularity conditions, asymptotic normality of the sequence of estimators generated by this approach are established. Usefulness of this approach in terms of saving computational time and achieving greater efficiency of estimation compared to the standard sequential approach are demonstrated with simulations conducted from two different sets of models.
\end{abstract}

\doublespacing

\section{Introduction} \label{sec:intro}

An extensively researched problem in design of experiments literature is to determine, under resource constraints, the optimal values of one or more controllable input variables $\mathbf{X}$ that would lead to the most efficient estimation of the parameters of a fitted specified model of a response $Y$ on $\mathbf{X}$. The general approach to finding such \emph{optimal designs} \citep{Atkinson2007} is to find $\mathbf{X}$ that optimizes a functional of the Fisher information matrix, such as its determinant or trace. This procedure is straightforward for linear models where the Fisher information matrix does not depend on the unknown model parameters. 
The problem, however, is much more complicated for situations where the Fisher information matrix depends on the unknown parameters, a phenomenon that is common in non-linear models such as standard nonlinear regression or generalized linear models. Consider, for example, the nonlinear exponential growth model considered in~\cite{LDHl2014}:
\begin{equation}
    y_i = \alpha_1 \exp \left( - \alpha_2 / x_i \right) + \epsilon_i, \  0 < x_{\min} < x_i < x_{\max}, i=1, \ldots, n, \label{eq:exponential}
\end{equation}
where for $i=1, \ldots, n$, $y_i$ and $x_i$ denote the observed values of $Y$ and $X$, and the residuals $\epsilon_i$'s are independently and identically distributed (iid) as $N(0, \sigma^2)$. As seen in \cite{LDHl2014}, the total Fisher information matrix for this model given the observed data points $(x_i, y_i)_{i=1, \ldots, n}$ is
\begin{equation}
    \mathbf{I}(\alpha_1, \alpha_2) = \frac{1}{\sigma^2} \sum_{i=1}^n \exp \left( - 2 \alpha_2 / x_i \right) \left[ 
    \begin{array}{cc} 
    1 & -\alpha_1/x_i \\
    -\alpha_1/x_i & \alpha_1^2/x_i^2 
    \end{array}
    \right]. \label{eq:M1Fisher}
\end{equation}

According to the criterion of D-optimality, the optimal solution to the problem of maximizing the determinant of the matrix in (\ref{eq:M1Fisher}) with respect to the input points $x_1, \ldots, x_n$ is known as the \emph{locally optimal design} \citep{Chernoff1953}. Clearly, the solution to this optimization problem depends on the unknown parameters ${\bm \theta} = (\alpha_1, \alpha_2)^\T$. \cite{Cochran1973} described this dependency: ``You tell me the value of ${\bm \theta}$, and I promise to design the best experiment for estimating ${\bm \theta}$''.

One way to get around this problem is to maximize the determinant of the Fisher information with respect to $\mathbf{x}$ at some ``guessed'' value or initial estimate of ${\bm \theta} = {\bm \theta}_0$ \citep[e.g.,][]{BoxLucas1959}. An obvious drawback of this approach, as noted by several authors is that, such a prior estimate ${\bm \theta}_0$ may be far from the true value of ${\bm \theta}$ and the behavior of the locally optimal design may be quite sensitive to even small perturbations in the parameter value. An alternative way to tackle this problem is to use a Bayesian approach \citep[e.g.][]{Chaloner1989} in which a suitable prior distribution is postulated for ${\bm \theta}$, and the expected value of the determinant of the Fisher information with respect to that prior is maximized to find the D-optimal design.

A more natural approach is to follow a sequential strategy consisting of two parts - an initial static design from which initial parameter estimates are obtained, followed by a fully sequential design where at every step the criterion function is optimized to find a new design point, and the parameter estimate is updated after observing the response at the new design point. Such a strategy was formalized by \cite{ChauMyk1993}, who proposed selecting design points sequentially using a D-optimality criterion, and showed that such a strategy leads to convergence of the design to the true D-optimal design under regularity conditions. Implementation of this procedure requires two optimization algorithms to be applied at each step - one to optimize the D-optimality criterion and the other to obtain the maximum likelihood estimate of the parameters. 

On the other hand, in many problems involving derivation of optimal designs for non-linear models, researchers are able to come up with analytical, closed-form solutions of such optimal designs, typically represented as a probability measure on the experiment space. For example, as shown by \cite{LDHl2014}, maximization of the Fisher information matrix in (\ref{eq:M1Fisher}) leads to a two-point balanced D-optimal design with equal support at points 
\begin{equation}
     x_1^* = \max \left( \frac{ \alpha_2 x_{\max}}{\alpha_2 + x_{\max}}, x_{\min} \right), x_2^* = x_{\max}. \label{eq:M1optdesign}
\end{equation}
Because such optimal designs depend on the unknown values of the parameters to be estimated, they cannot be utilized to their fullest potential and often remain of academic interest only.

The work in this paper is motivated by the intuition that incorporating such existing analytical solutions into a sequential algorithm would create a win-win situation for both the approaches. On one hand, it would absolve the experimenter from optimizing the criterion function at each step leading to significant saving of computational time for complex models with a large number of parameters. On the other hand, it would effectively leverage the analytical solutions obtained through tedious mathematical derivations.

In this paper, we propose such a hybrid sequential strategy that replaces the optimization of the objective function at every single step by \emph{plugging in} the estimates into the available closed form solutions, and generating new design points as draws from the probability distributions induced by the optimal design. We refer to this sequential strategy as PICS (Plug into closed-form solution). We provide theoretical guarantees for the proposed approach by establishing that the key results derived by \cite{ChauMyk1993}, i.e., convergence of the sequence of design points to the true D-optimal design and asymptotic normality of the sequence of parameter estimates, also hold in the PICS approach.

The paper is organized as follows: in Section \ref{sec:prelim}, we introduce key notations, briefly review D-optimal designs and provide two specific examples of our setting. We present the two stage approach proposed by \cite{ChauMyk1993} in Section \ref{sec:sequential}. In Section \ref{sec:PICS}, we describe the proposed PICS approach, and state the key theoretical results. In Section \ref{sec:Simulations} we re-visit the motivating examples presented in Section \ref{sec:prelim}, and demonstrate how the proposed approach outperforms the standard approach using extensive simulations. The final section presents some concluding remarks and directions for future work.

\section{Background, notation, problem set up and examples}\label{sec:prelim}
%

Our set up is similar to the one in \cite{ChauMyk1993}, where the aim is to fit a model governing the non-linear relationship between a response variable $Y$ and a deterministic $p$-dimensional vector of regressors ${\mathbf X}$ using data $(y_i, \mathbf{x}_i)$, $i=1, \ldots, n$, obtained from a controlled experiment. The response $Y$ takes values from the response space $\mathbf{\Rcal}$. The regressor $\mathbf{X}$ takes values chosen by the experimenter from an experiment space $\exptspace$, which can be a finite set, a subset of the Euclidean space, or a combination of both depending on whether $\mathbf{X}$ is categorical, continuous, or a combination of both types of variables. The conditional distribution of $Y$ given $\mathbf{X} = \mathbf{x}$ will be denoted by $f(y |\bm{\theta}, \phi, \mathbf{x})$, assumed to be completely known except for the values of $\bm{\theta}$ and $\phi$. The parameter $\bm{\theta} \in \bm{\Theta}$ is the parameter of interest, and $\phi$ is a nuisance parameter. As in \cite{ChauMyk1993}, we will assume that the nuisance parameter $\phi$ has no influence on the computation of the maximum likelihood estimate (MLE) of $\bm{\theta}$, and hence for simplicity of notation, will write our model as $f(y| \bm{\theta}, \mathbf{x})$. The Fisher information matrix at $\trueparm$ is defined as:
$$ \mathbf{I}(\bm{\theta}, \mathbf{x}) =\int_{\Rcal} \score \{\score\}^{\T} \likelihood \mu (dy),$$ where $\nabla$ is the gradient operator denoting differentiation with respect to $\anytrueparm \in \bm{\Theta}$, and $\mu$ denotes the standard Lebesgue or counting measure with respect to $\Rcal$.

Following \cite{Kiefer1959, Kiefer1961a} and \cite{Kiefer_Wolfowitz}, the collection of all probability measures on $\exptspace$ is called the \emph{design space}, and is denoted by $\designspace$. A specific design $\anydesign \in \designspace$ is a probability measure on $\exptspace$. A D-optimal design \citep{Atkinson2007} is obtained by maximizing the determinant of $\mathbf{I}(\bm{\theta}, \mathbf{x})$ over all possible designs $\anydesign \in \designspace$. Based on the definition of ``locally optimal'' design in \cite{Chernoff1953}, a locally D-optimal design is defined as follows:
\begin{definition}\label{def}
A design $\optdesign \in \designspace$ is called a \emph{locally D-optimal design}  at $\trueparm$ if $\det  \left( \int_{\exptspace} \mathbf{I}(\bm{\theta}, \mathbf{x}) \optdesign ( d\mathbf{x} ) \right) = \sup_{\anydesign \in \designspace} \det  \left( \int_{\exptspace} \mathbf{I}(\bm{\theta}, \mathbf{x}) \anydesign ( d\mathbf{x} ) \right). $
\end{definition}
As pointed out by \cite{ChauMyk1993}, such a $\optdesign$ will always exist in an application, as $\exptspace$ will typically be a finite set or some nice subset of an Euclidean space. However a closed-form solution to this optimization problem may not exist. \cite{ChauMyk1993} gave several examples of the setting described above. We consider cases where $\optdesign$ is available in the form of a known probability distribution over $\exptspace$ that is completely known if $\bm{\theta}$ is known, so that for a given value of $\trueparm$, it is possible to generate a draw $\mathbf{X}^*(\trueparm)$ from the distribution of the locally optimal design $\optdesign$. We provide two specific examples of such a setting to motivate our approach and to demonstrate its effectiveness and advantage.

\begin{example}[Non-linear regression] \label{example1}
  Consider a non-linear regression model where $y$ and $\mathbf{x}$ are related via the equation
 \begin{equation}
     y_i = g(\trueparm, \mathbf{x}_i) + \epsilon_i, \ i=1, \ldots, n, \label{eq:nonlinearreg}
 \end{equation} 
where $g(\cdot)$ is any real-valued non-linear function with a known form, and the random errors $\epsilon_1, \ldots, \epsilon_n$  are iid normal variables with zero mean and unknown variance $\sigma^2$. A specific example is (\ref{eq:exponential}), where $\mathbf{x}$ is a scalar, $g(\trueparm, \mathbf{x}) = \alpha_1 \exp(-\alpha_2 / x)$ and the experiment space $\exptspace$ is the closed interval $[\Xmin, \Xmax]$. The vector $(\alpha_1, \alpha_2)$ is the parameter $\trueparm$ of interest and $\sigma^2$ is the nuisance parameter. As mentioned in Section \ref{sec:intro}, the locally D-optimal design $\optdesign$ can be expressed as a known probability distribution on $\exptspace$, which for model (\ref{eq:exponential}) is a balanced design with equal support at the two points given in (\ref{eq:M1optdesign}).

\cite{LDHl2014} also considered other forms of $g(\trueparm, \mathbf{x})$, and derived closed form expressions for the optimal designs under each model, the details of which will be provided in Section \ref{sec:Simulations}.

\end{example}

\begin{example}[Logistic regression in factorial experiment] \label{example2}
Consider a $2^2$ factorial experiment in which the response is binary and each of the two input variables $X_1$ and $X_2$ can take two possible coded levels $+1$ and $-1$, so that the experiment space consists of four possible level combinations $(+1,+1), (+1,-1), (-1,+1)$ and $(-1,-1)$. We assume that the conditional distribution of $Y$ given $X_1 = x_1$ and $X_2 = x_2$ is Bernoulli with parameter 
\begin{equation}
\pi(x_1, x_2) = \frac{\exp(\beta_0 + \beta_1 x_1 + \beta_2 x_2)}{1 + \exp(\beta_0 + \beta_1 x_1 + \beta_2 x_2)}, \label{eq:logistic}
\end{equation}
where
${\bm \theta} = (\beta_0, \beta_1, \beta_2)^\T$ is the parameter to be estimated. 

In this example, a design $\anydesign$ can be defined in terms of the proportion of units $p_{rs}$ allocated to the $r$th level of $x_1$ and the $s$th level of $x_2$, for $r,s=1,2$. For example, $p_{11}$ represents the proportion of units allocated to level combination $(+1,+1)$. These proportions $p_{rs}$ depend on the unknown parameter $\trueparm$. Under specific assumptions, \cite{Mandal2012} derived closed form solutions for the D-optimal design $\optdesign$ as the proportions $p_{rs}^*$. The explicit forms of these optimal designs will be provided in Section \ref{sec:Simulations}.
\end{example}

\section{The two-stage approach} \label{sec:sequential}

\cite{ChauMyk1993} proposed a two-stage sequential design strategy in order to resolve the awkward situation encountered in non-linear models, where  designing an efficient experiment requires knowledge of the parameter, but the purpose of the experiment is to generate data to yield parameter estimates. We will refer to this approach as the C-M approach throughout the remainder of this paper.

The C-M procedure, described algorithmically in Figure~\ref{fig:algCM}, assumes a predetermined design size $n$ based on resource constraints, and divides the process of generating these $n$ design points into the following two stages: 
\begin{enumerate}[label=(\arabic*)]
    \item an initial \emph{static} stage consisting of $\initialrun$ trials, \label{cma}
    \item a \emph{fully adaptive sequential} stage consisting of the remaining $\designsize-\initialrun$ trials. \label{cmb}
\end{enumerate}
In stage~\ref{cma}, the $n_1$  design points are chosen with minimal prior knowledge on $\trueparm$. The initial design points $\mathbf{X}_1, \ldots, \mathbf{X}_{n_1}$ are drawn from an initial distribution $\initialdesign$. For example, $\initialdesign$ can be a uniform distribution over $\exptspace$, or can constitute a Latin hypercube sample \citep{Mackay_LHD} from $\exptspace$. Let $Y_1, \ldots, Y_{n_1}$ be the responses observed after the initial experiment is carried out. The goal of this stage is to find a starting estimate of $\trueparm$ based on the data $\left(Y_j, \mathbf{X}_j \right), j= 1, 2, \ldots, \initialrun$, observed from the initial experiment. One can use the method of maximum likelihood to obtain an estimate $\mle{\initialrun}$ of $\trueparm$. 

This estimate $\mle{\initialrun}$ is used to initiate the sequential stage~\ref{cmb}, where the experimenter finds the next design point and the estimate of $\trueparm$ at each iteration. Prior to the $i^{th}$ iteration, where $\initialrun+1 \leq i \leq \designsize$, the data $ \left( Y_j, \mathbf{X}_j\right), j= 1, 2, \ldots, i-1$ and the ML estimate $\mle{i-1}$ of $\trueparm$ based on this data are available. The subsequent design point $\mathbf{X}_i \in \exptspace$ is chosen by maximizing the criterion function 
\begin{equation}\label{criterion}
    \det \left [ \sum_{j=1}^{i-1} \mathbf{I} \left( \mle{i-1}, \mathbf{X}_j \right) + \mathbf{I}\left(\mle{i-1}, \mathbf{X}_i \right) \right ],
\end{equation}
with respect to $\mathbf{X}_i$. Note that the criterion function represents the determinant of the total Fisher information matrix  $\sum_{j=1}^i \mathbf{I} \left(\trueparm, \mathbf{X}_j \right)$ till stage $i$.

Next, the response $Y_i$ at $\mathbf{X}_i$ is observed, and the estimate of $\trueparm$ is updated based on $ \left(Y_j, \mathbf{X}_j \right), j= 1, 2, \ldots, i$. The updated estimate is denoted as $\mle{i}$. These actions are repeated until a set of $\designsize$ design points are obtained.

\begin{figure}[!ht]
		\caption{The C-M Algorithm} \label{fig:algCM}
		\centering
		\fbox{
			\begin{minipage}{12 cm}
				\begin{enumerate}
					\item Stage 1 (Static stage):
                        \begin{enumerate}
                            \item Choose the initial design points $\mathbf{X}_1, \ldots, \mathbf{X}_{n_1} \drawiid \initialdesign$ on $\exptspace$, where $\initialdesign$ is an initial probability measure.
                            \item Observe the corresponding set of responses $Y_1, \ldots, Y_{n_1}$.
                            \item Calculate the ML estimate $\mle{\initialrun}$ based on the datapoints $\left(Y_j, \mathbf{X}_j \right)$, $j= 1, 2, \ldots, \initialrun$.
                        \end{enumerate}
					 \item Stage 2 (Sequential stage):
					for {$\initialrun+1 \leq i \leq \designsize$,} 
                      \begin{enumerate}
    \item Given $\mle{i-1}$ and $ \left(Y_j, \mathbf{X}_j \right), j= 1, 2, \ldots, i-1$, choose
        \begin{equation*}
        \mathbf{X}_i = \arg\max_{\mathbf{X}_i \in \exptspace}  \det \left [ \sum_{j=1}^{i-1} \mathbf{I} \left( \mle{i-1}, \mathbf{X}_j \right) + \mathbf{I}\left(\mle{i-1}, \mathbf{X}_i \right) \right ]
        \end{equation*}
    \item Observe response $Y_i$ at $\mathbf{X}_i$.
    \item Update the ML estimate $\mle{i}$ based on the data $ \left(Y_j, \mathbf{X}_j \right), j= 1, 2, \ldots, i$.
    \end{enumerate}
					
				\end{enumerate}
			\end{minipage}
   }
	\end{figure}

A key aspect of the sequential procedure described above is that, it leads to a dependent structure of the data. As a result, the total Fisher information $\sum_{j=1}^{n} \mathbf{I} \left( \trueparm, \mathbf{X}_j \right)$ becomes a random variable. \cite{ChauMyk1993} established the following asymptotic results under this setting:
\begin{enumerate}
    \item The maximum likelihood estimator $\mle{\designsize}$ of $\trueparm$ based on the data $(Y_j, \mathbf{X}_j), j= 1, 2, \ldots,\designsize$, is shown to be consistent and asymptotically normal and efficient \cite[Theorem 3.10 and Corollary 3.11]{ChauMyk1993}. 
    \item The generated sequential design is shown to converge to the true locally D-optimal design at $\trueparm$ in probability \cite[Theorem 3.5]{ChauMyk1993} when $\initialrun \rightarrow \infty$ and $\frac{\initialrun}{\designsize} \rightarrow 0$  as $\designsize \rightarrow \infty$. 
\end{enumerate}

However, it is worthwhile to note that this method involves \emph{two} primary optimization algorithms at each step: (i) determining the subsequent design point by optimizing equation~(\ref{criterion}) and (ii) updating the parameter estimates from the augmented data by maximizing the likelihood. The computational burden associated with (i) can be prohibitive, especially as the model becomes more complex and the dimension of the input space increases. We aim at harnessing existing analytical results to eliminate the first optimization and reduce the computational effort, while retaining the nice asymptotic properties of the C-M procedure.

\section{The sequential PICS design: algorithm and theoretical guarantees} \label{sec:PICS}

We now describe our proposed approach, the PICS algorithm, which is very similar to the C-M procedure discussed in Section \ref{sec:sequential} but takes advantage of closed-form locally D-optimal designs when they are available. To the best of our knowledge, the significant amount of research that has led to identification of closed-form solutions of optimal designs for non-linear models have primarily been used to identify locally optimal designs by substituting estimates or ``guestimates'' of the parameters into such solutions. Some researchers have taken the Bayesian route by postulating prior distributions for the parameters \citep[e.g.,][]{LDHl2014}. Whereas the Bayesian approach permits the experimenter to use prior knowledge about the parameter, it makes the optimization problem more complicated than its sequential counterpart. We propose to use such results to effectively eliminate the computationally expensive step (a) of Stage 2 of Figure~\ref{fig:algCM} by plugging in the parameter estimates obtained from the previous stage directly into the closed-form solution, and consequently boost the computational efficiency of the C-M procedure while maintaining its nice asymptotic properties. The PICS algorithm, shown in Figure~\ref{fig:algPICS}, is thus essentially the same as the C-M algorithm summarized in Figure~\ref{fig:algCM}, except for the first step of Stage 2, where instead of optimizing over the criterion function~(\ref{criterion}), the new design point $\mathbf{X}_i$ is generated as a draw from the known optimal design measure. 

\begin{figure}[!ht]
		\caption{The PICS Algorithm} \label{fig:algPICS}
		\centering
		\fbox{
			\begin{minipage}{12 cm}
				\begin{enumerate}
                        \item Stage 1 (Static stage):
                        \begin{enumerate}
                            \item Choose the initial design points $\mathbf{X}_1, \ldots, \mathbf{X}_{n_1} \drawiid \initialdesign$ on $\exptspace$, where $\initialdesign$ is an initial probability measure.
                            \item Observe the corresponding set of responses $Y_1, \ldots, Y_{n_1}$.
                            \item Calculate the ML estimate $\mle{\initialrun}$ based on the datapoints $\left(Y_j, \mathbf{X}_j \right)$, $j= 1, 2, \ldots, \initialrun$.
                            \end{enumerate}
					 \item Stage 2 (Sequential stage):
					for {$\initialrun+1 \leq i \leq \designsize$,} 
                      \begin{enumerate}
    \item Given $\mle{i-1}$, choose $\mathbf{X}_i$ by generating a draw from the optimal design $\optdesign(\mle{i-1})$, i.e., by substituting $\trueparm = \mle{i-1}$ in the probability distribution associated with the closed-form optimal design.
    \item Observe response $Y_i$ at $\mathbf{X}_i$.
    \item Update the ML estimate $\mle{i}$ based on the data $ \left(Y_j, \mathbf{X}_j \right), j= 1, 2, \ldots, i$.
    \end{enumerate}
					
				\end{enumerate}
			\end{minipage}
   }
	\end{figure}

Thus replacing the optimization step involved in obtaining the next design point by a draw from the known optimal design measure is the only difference between the C-M and PICS approaches. Although drawing from a known distribution is straightforward, implementation of a \emph{fully} sequential design (i.e., generating \emph{one} new design point at every step) involves some nuances. Consider, for example, the optimal design for Example \ref{example1} given in~(\ref{eq:M1optdesign}) with equal support at $X^*_1$ and $X^*_2$. At each step, if we select one of the two points with probability $1/2$, for a small $n$, the design can be heavily imbalanced with a moderate probability. Such possibilities, however, can be handled by slightly adjusting the drawing procedure associated with step (a) of Stage 2 in the PICS algorithm shown in Figure \ref{fig:algPICS}. For example, if $\optdesign$ is a balanced design with $k$ unique support points at $\mathbf{X}^*_1(\trueparm), \ldots, \mathbf{X}^*_k(\trueparm)$, instead of a fully sequential algorithm, it will be pragmatic to use a batch-sequential algorithm that generates $k$ new points at each iteration in the algorithm with loops of $\blncdesignsize$ iterations, where in each loop each of the support points is chosen once and exactly once, in a random order. This procedure can also be used to accommodate unbalanced designs with finite supports. For example, if the optimal design has support at three points $\mathbf{X}_1^*$, $\mathbf{X}_2^*$ and $\mathbf{X}_3^*$ with probabilities 1/4, 1/2 and 1/4 respectively, then one can use a batch-sequential algorithm as above with a batch size of four by re-expressing the optimal design as a four-point balanced design with equal support at $\mathbf{X}_1^*, \mathbf{X}^*_2, \mathbf{X}_2^*$ and $\mathbf{X}_3^*$. We will refer this batch sequential procedure as the \emph{balanced PICS algorithm} because it enforces the exact balance of the true optimal design into the PICS procedure.

Like the C-M approach, we assume that for the PICS approach, the total number of experimental trials $n$ that can be conducted is fixed and determined apriori by resource constraints. However, in more flexible situations, one can also use a stopping criterion based on the relative improvement achieved with the D-optimality criterion in comparison to the previous step, i.e., 
$$ \left \lvert \frac{\det \left (\sum_{j=1}^i \mathbf{I}( \widehat{\trueparm}_i, \mathbf{X}_j) \right ) - \det \left (\sum_{j=1}^{i-1} \mathbf{I}( \widehat{\trueparm}_{i-1}, \mathbf{X}_j) \right ) }{\det \left ( \sum_{j=1}^{i-1} \mathbf{I}( \widehat{\trueparm}_{i-1}, \mathbf{X}_j)\right ) } \right \rvert < \delta,$$
where $\delta$ is a pre-assigned threshold.

\subsection{Theoretical guarantees of the PICS algorithm} \label{ss:mainresults}

In this section, we establish theoretical results for the proposed PICS algorithm summarized in Figure~\ref{fig:algPICS} that are similar to those guaranteed by the C-M algorithm. We first establish a Theorem that guarantees that the observed average Fisher information matrix calculated based on the design points obtained using the PICS approach will converge in probability to the true expected Fisher information matrix 
\begin{equation}
\mathbf{I}^* = \int \mathbf{I} (\trueparm, \mathbf{x}) \optdesign (d \mathbf{x}) \label{eq:trueFI}
\end{equation}
at the true optimal design $\optdesign$ as the number of trials goes to infinity.

\begin{theorem} \label{theorem1} (Consistency of the optimal design)
    Under regularity conditions (A.1)-(A.3), for any $\trueparm \in \bm{\Theta}$, the design points $\mathbf{X}_1, \ldots, \mathbf{X}_n$  chosen using the PICS approach satisfies the following property:
\begin{equation}
    \frac{\sum_{i=1}^n \mathbf{I}(\trueparm, \mathbf{X}_i)}{n} \convergep \mathbf{I}^* \quad \text{as} \quad \designsize \rightarrow \infty. \label{eq:theorem1_1} 
    \end{equation} 
\noindent Additionally, if condition (A.4) is satisfied, then
    \begin{equation}
    \frac{\sum_{i=1}^n \mathbf{I}(\widehat{\bm{\theta}}_{n}, \mathbf{X}_i)}{n} \convergep \mathbf{I}^* \quad \text{as} \quad \designsize \rightarrow \infty. \label{eq:theorem1_2}
    \end{equation} 
\end{theorem}

The proof of this theorem can be found in Section~\ref{sec:Proof-of-thm-1}. 
Next, we prove the asymptotic normality of the sequence of maximum likelihood estimators $\{\mle{n}\}$ of the true parameter $\trueparm$ based on a set of regularity conditions.

\begin{theorem}\label{theorem2} Under regularity conditions (A.1)-(A.8), the maximum likelihood estimator $\mle{n}$ of $\trueparm$ satisfies
\begin{equation*}
    \sqrt{\designsize} (\mle{n} - \trueparm) \converged \mathcal{N}_\dimn \left(0, \istarinv\right)
\end{equation*}
where $\istar$ is the true optimal Fisher information defined in (\ref{eq:trueFI}). Additionally, invoking Theorem~\ref{theorem1} we deduce
\begin{equation*}
        \left\{\sum_{i=1}^n \mathbf{I}(\widehat{\bm{\theta}}_{n}, \mathbf{X}_i) \right\}^{\frac{1}{2}}\left(\mle{n} - \trueparm\right) \converged \mathcal{N}_\dimn(0, \mathrm{I}_\dimn)
    \end{equation*}
\end{theorem} 
See Section~\ref{sec:Proof-of-thm-2} for a proof of this theorem. A few remarks on Theorems~\ref{theorem1} and~\ref{theorem2} are given below. 

\begin{remark}\label{remark1}
We detail the regularity conditions~\ref{condition:A.1}-\ref{condition:A.8} as referred to in the statements of Theorem~\ref{theorem1} and Theorem~\ref{theorem2} in Section~\ref{sec:regularity-conds}. Conditions~\ref{condition:A.1}-\ref{Condition:A.3} impose a few smoothness assumptions (Lipschitz property) on the Fisher information matrix $\mathrm{\bf I}(\trueparm, \bf X)$. These assumptions are mild and hold for most non-linear models.
In Assumption~\ref{Condition:A.4}, we posit that the sequence of maximum likelihood estimators $\{\mle{n}\}$ obtained from the PICS method is strongly consistent. For generalized linear models, the strong consistency of $\{\mle{n}\}$ for sequentially collected data, similar to PICS, is guaranteed under a very mild assumption~\citep{Chen1999}. In particular, if the number of initial design points $n_1$ satisfies $n_1^{-1} (\log n) \rightarrow 0$ as $n \rightarrow \infty$, the design points are uniformly bounded, and the initial design points are chosen at random, then this strong assumption is satisfied. For nonlinear models, a similar condition can be found in the works of~\cite{lai1994asymptotic}.Conditions~\ref{condition:A.5}-\ref{condition:A.8} impose smoothness conditions on the log-likelihood function and are analogous to the smoothness conditions from~\cite{ChauMyk1993}.
\end{remark}

\begin{remark}\label{remark2}
Theorem~\ref{theorem2} provides assurance that the ML estimate $\mle{n}$ of $\trueparm$ maintains asymptotic efficiency, akin to the conventional maximum likelihood estimate derived from independent observations. Furthermore, the theorem implies that inference procedures for $\trueparm$ remain robust even in scenarios where the design is generated sequentially. Theorem~\ref{theorem2} also allows us to construct confidence ellipsoids for $\trueparm$ using the maximum likelihood estimate $\mle{n}$ and the average Fisher information matrix. 
\end{remark}

\begin{remark} \label{remark:consistency-of-design}
Theorem~\ref{theorem2} ensures that our sequential estimates $\{ \mle{i} \}_{i \geq 1}$ obtained during the PICS-strategy are asymptotically efficient (optimal limiting variance) for the parameter $\trueparm$. We can, in fact, provide a similar guarantee on the sequence of designs $\left\{\optdesign{(\mle{i})} \right\}_{i \geq 1}$ that are used to draw samples at each round of the PICS algorithm. Let $\optdesign$ be the optimal design, and we assume that $\trueparm \mapsto \varphi(\optdesign{(\trueparm)})$ --- the map from a point $\trueparm$ to the characteristic function of the design $\optdesign{(\trueparm)}$ --- is a continuous function of $\trueparm$. Then, using the strong consistency of  $\{ \mle{i} \}_{i \geq 1}$, we have 
\begin{align*}
    \optdesign{(\mle{n})} \stackrel{d}{\rightarrow} \optdesign 
    \qquad \text{as} \;\; n \rightarrow \infty.
\end{align*}  
Put simply, the estimated designs used in the PICS algorithm themselves converge to the optimal design. Figure~\ref{fig:densityM3} provides a numerical validation of this convergence of the estimated design to the optimal design. 
\end{remark}

\begin{remark}  \label{remark:Martigale-clt}
We note that the primary challenge in proving the above results stems from the dependency among the sequence of points. Hence, unlike using the concepts of the usual Central Limit Theory, we must draw upon ideas rooted in martingale limit theory to derive these results. 
\end{remark}

\section{Simulations from real experiment settings} \label{sec:Simulations}

In this section, we conduct simulations with the motivating examples described earlier in Section \ref{sec:prelim} to demonstrate how the PICS procedure works, and to compare its performance with the C-M algorithm with respect to statistical and computing efficiency.

\subsection{Example 1: Non-linear regression} \label{ss:simul_nlr}

 We consider three non-linear models $M_1, M_2$ and $M_3$ discussed by \cite{LDHl2014} that can describe the growth pattern of nanostructures over time (input variable $X$) in minutes. The response variable $Y = Y(x)$ represents the length of nanostructures when grown till time $X=x$. The experiment space $\exptspace$ is the closed interval $\lbrack \Xmin = 0.5 , \Xmax = 210 \rbrack$. Model $M_1$ represents the pure exponential growth model (\ref{eq:exponential}) already introduced in Section \ref{sec:intro}.  The second model $M_2$ is characterized by an exponential-linear change-point growth function given by:
\begin{equation} \label{eq:M2}
    \growthfn{\inps} = \begin{cases}
                        \gexpo, & \Xmin \leq x < \changept\\
                        \glin, & \changept \leq x \leq \Xmax
                        \end{cases}
\end{equation}
\noindent where $\alphaone$, $\alphatwo$ and $\slope$ are all greater than zero, and $\changept$ is the \textit{known} change point. The third model $M_3$ differs from $M_2$ by the fact that the change point $\changept$ is an unknown parameter in $M_3$. Hence, both models have four common parameters of interest, namely $\alphaone$, $\alphatwo$, $\intercept$ and $\slope$, and a common nuisance parameter $\sigma^2$. Model $M_3$ contains an additional parameter $\changept$ of interest. Assuming continuity of $\growthfn{\inps}$ and its first derivative $g^{\prime}(\bm{\theta}, \mathbf{x})$ at $\changept$,  models $M_2$ and $M_3$ reduce to two- and three-parameter models respectively \cite[Section 2]{LDHl2014}, so that model $M_2$, like $M_1$, can be parameterized by $\trueparm = ( \alphaone, \alphatwo )^\T$, whereas for $M_3$, $\trueparm = ( \alphaone, \alphatwo, \changept )^\T$ is the parameter of interest. In the simulation setup, the estimated values of the parameters reported from \cite{LDHl2014} as $\modeltrueparmval{\alpha}{1} = 32.11$, $\modeltrueparmval{\alpha}{2} = 105.65$, $\changept^* = 86.67$ and $\modeltrueparmval{\sigma^2}{} = 0.086$ are considered true parameter values and used to simulate the response $Y$ for given values of $X$.

To implement the C-M approach for each model using the algorithm described in Figure \ref{fig:algCM}, we first simulate an initial design of $n_1$ runs, generate the response values and compute the MLE of $\trueparm$ from these observations by maximizing the likelihood using the minimize function in python. Next, we move to the adaptive stage where subsequent design points are obtained by maximizing (\ref{criterion}), a new response is generated and the MLE is updated. To execute this step, we use the expressions for the Fisher information matrix for the particular model from \cite{LDHl2014}.

To implement the PICS approach using the algorithm described in Figure \ref{fig:algPICS}, the initial static design is generated in the same manner as above. In the adaptive stage, we obtain the subsequent design points by drawing an observation from the probability distribution associated with the following locally D-optimal designs for models $M_1, M_2$ and $M_3$ established by \cite{LDHl2014}, plugging-in the MLE of $\trueparm$ obtained from the previous step: 
\begin{enumerate}
    \item $M_1$: A balanced two-point design with unique support at points $\optpts{\trueparm}{1}$ and $\optpts{\trueparm}{2}$, where 
    $$\optpts{\trueparm}{1} = \max \left( \frac{\alphatwo\optpts{\trueparm}{2}}{\alphatwo + \optpts{\trueparm}{2}} , \Xmin \right), \  \optpts{\trueparm}{2} = \Xmax.$$
    \item $M_2$: A balanced two-point design with unique support at points $\optpts{\trueparm}{1}$ and $\optpts{\trueparm}{2}$, where 
    $$\optpts{\trueparm}{1} = \max ( \tau , \Xmin), \optpts{\trueparm}{2} = \Xmax,$$
    where $$\tau = \frac{\alphatwo}{ 1 - \alphatwo \frac{( \Xmax - 2 \changept ) \changept - \alphatwo ( \Xmax - \changept )}{ \changept ( \changept^2 + \alphatwo   ( \Xmax - \changept ))}}.$$
    \item $M_3$: A balanced three-point design with unique support at points $\optpts{\trueparm}{1}$, $\optpts{\trueparm}{2}$ and $\optpts{\trueparm}{3}$, where 
    $$\optpts{\trueparm}{1} = \max \left( \frac{\alphatwo\changept}{\alphatwo + \changept} , \Xmin \right), \ \optpts{\trueparm}{2} = \changept , \ \optpts{\trueparm}{3} = \Xmax.$$
\end{enumerate}

We implement and compare two versions of the PICS approach - the fully sequential algorithm shown in Figure \ref{fig:algPICS}, and balanced PICS, the balanced sequential algorithm where designs with $k$ support points are generated in loops of $k$, generating each point once and only once within each loop. Both these approaches are compared to the C-M approach using the following visualization tools and criteria:

\begin{enumerate}
\item Density plots of all points generated during the sequential stage.
\item The relative efficiency at step $i = 1, \ldots, n$, defined as
\begin{equation*}
    e_i = 1 - \left | \frac{ \det \left(\frac{1}{i} \sum_{j=1}^i \mathbf{I} (\mle{i}, \mathbf{X}_j ) \right) - \det(\mathbf{I}^* )}{\det (\mathbf{I}^*)} \right |,
\end{equation*}
    where $\mathbf{X}_j$ is the design point chosen at step $j$, and $\mathbf{I}^*$ given by (\ref{eq:trueFI}) is the true Fisher information matrix at the true value of the parameter, known for each model. Higher the $e_i$, closer is the estimated Fisher information till stage $i$ to the true Fisher information. Plotting $e_i$ against $i=1, \ldots, n$ provides a visual representation of the convergence of Theorem \ref{theorem1}.
\item The computing time required to generate $n$ design points.   
\end{enumerate}

\begin{figure}[htp]
\centering
  \begin{subfigure}{0.5\textwidth}
    \centering
    \includegraphics[width=\linewidth]{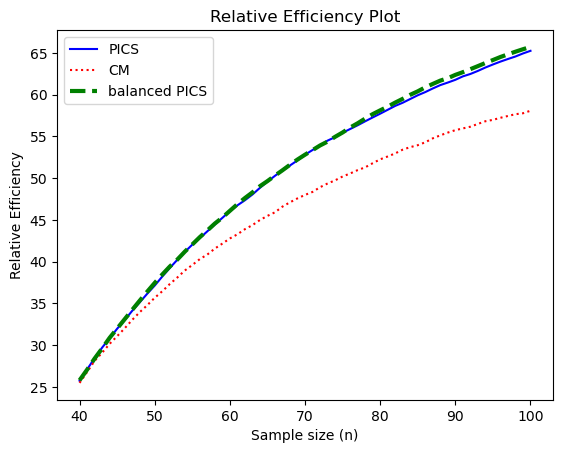}
    \caption{$n_1 = 40, n = 100$, 3-point initial design}
    \label{fig:subfig1ia}
  \end{subfigure}%
  \begin{subfigure}{0.5\textwidth}
    \centering
    \includegraphics[width=\linewidth]{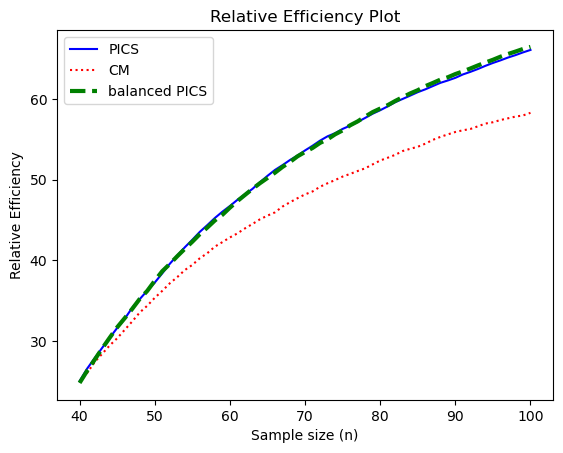}
    \caption{$n_1 = 40, n = 100$, uniform initial design}
    \label{fig:subfig1ib}
  \end{subfigure} \\
    \begin{subfigure}{0.5\textwidth}
    \centering
    \includegraphics[width=\linewidth]{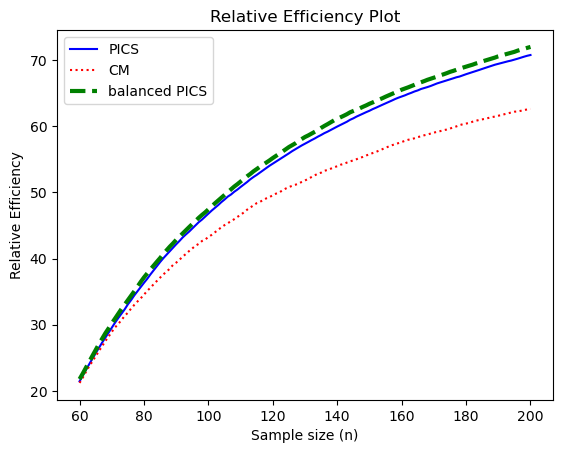}
    \caption{$n_1 = 60, n = 200$, 3-point initial design}
    \label{fig:subfig1iia}
  \end{subfigure}%
  \begin{subfigure}{0.5\textwidth}
    \centering
    \includegraphics[width=\linewidth]{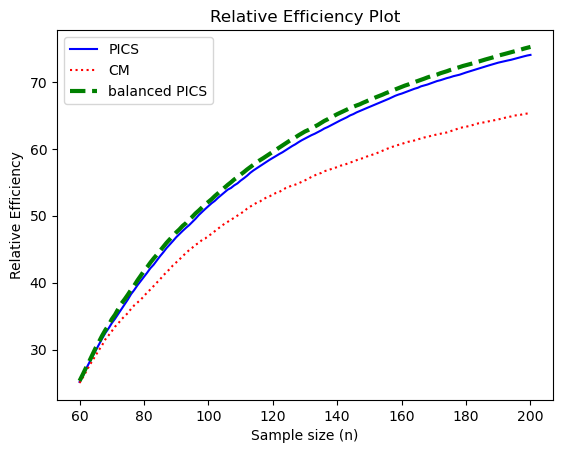}
    \caption{$n_1 = 60, n = 200$, uniform initial design}
    \label{fig:subfig1iib}
  \end{subfigure}
     \caption{Comparison of the three procedures with respect to efficiency for $M_2$} 
    \label{fig:efficiencyM2}
\end{figure}

\begin{figure}[ht]
\centering
  \begin{subfigure}{0.5\textwidth}
    \centering
    \includegraphics[width=\linewidth]{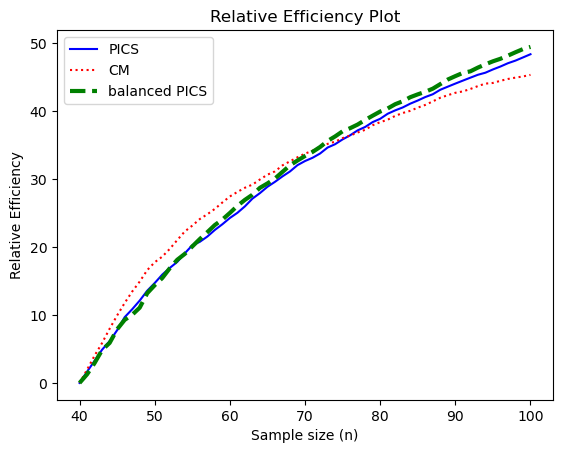}
    \caption{$n_1 = 40, n = 100$, 3-point initial design}
    \label{fig:subfig1iiia}
  \end{subfigure}%
  \begin{subfigure}{0.5\textwidth}
    \centering
    \includegraphics[width=\linewidth]{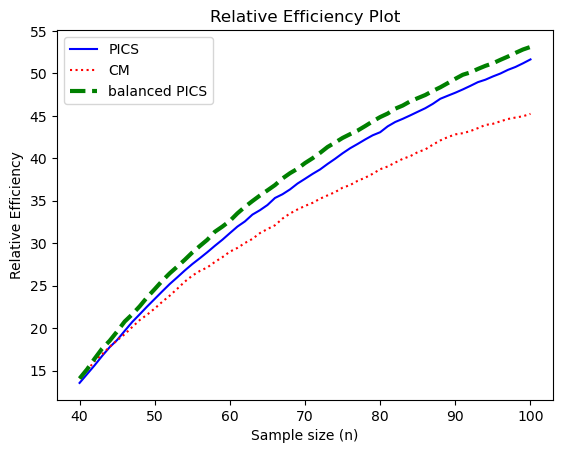}
    \caption{$n_1 = 40, n = 100$, uniform initial design}
    \label{fig:subfig1iiib}
  \end{subfigure} \\
    \begin{subfigure}{0.5\textwidth}
    \centering
    \includegraphics[width=\linewidth]{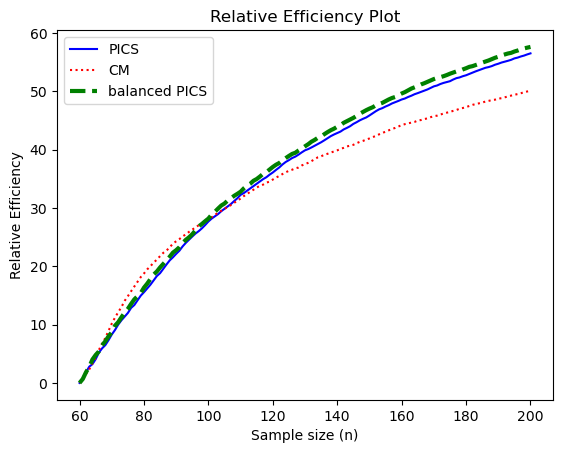}
    \caption{$n_1 = 60, n = 200$, 3-point initial design}
    \label{fig:subfig1iva}
  \end{subfigure}%
  \begin{subfigure}{0.5\textwidth}
    \centering
    \includegraphics[width=\linewidth]{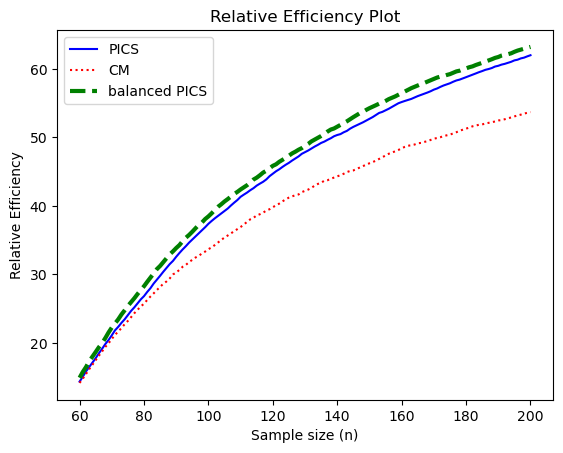}
    \caption{$n_1 = 60, n = 200$, uniform initial design}
    \label{fig:subfig1ivb}
  \end{subfigure}
     \caption{Comparison of the three procedures with respect to efficiency for $M_3$} 
    \label{fig:efficiencyM3}
\end{figure}

Two choices of the initial and final design size $(n_1, n)$: (i) ($n_1 = 40, n=100$) and (ii) ($n_1 = 60, n=200$); and two choices of $\xi_0$, the initial design - (i) uniformly distributed over $\exptspace$ and (ii) distributed at $\Xmin$, $\Xmax$ and $(\Xmin + \Xmax)/2$ with weights $(0.3,0.3,0.4)$ were considered. Each model was tested with 50 simulations under each setting. The efficiency curves shown in Figure \ref{fig:efficiencyM2} and Figure \ref{fig:efficiencyM3} and the computing times shown in Table \ref{table:comparisonnlm} are based on means of these 50 simulations. 


\begin{figure}[ht]
\centering
  \begin{subfigure}{0.45\textwidth}
    \centering
    \includegraphics[width=\linewidth]{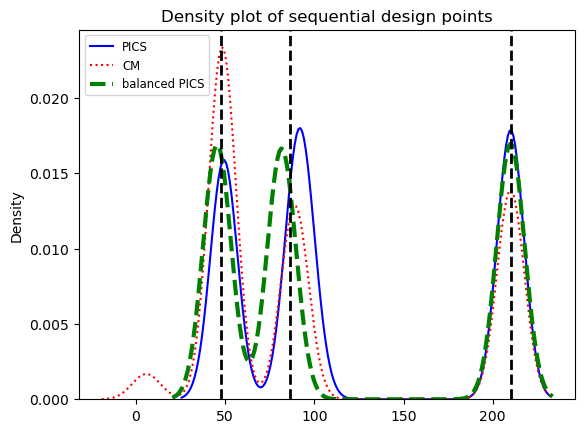}
    \caption{$n_1 = 40, n = 100$, uniform initial design}
    \label{fig:subfig1va}
  \end{subfigure}%
  \begin{subfigure}{0.45\textwidth}
    \centering
    \includegraphics[width=\linewidth]{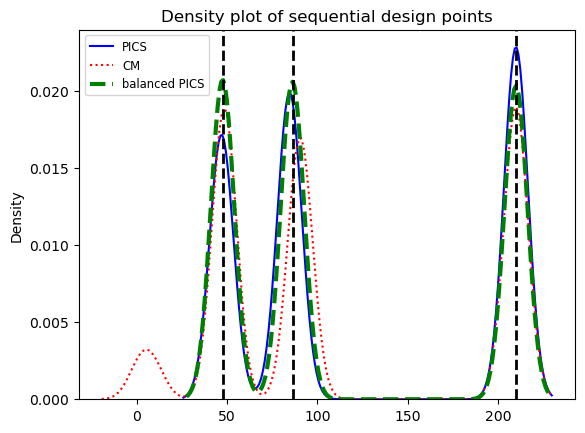}
    \caption{$n_1 = 60, n = 200$, uniform initial design}
    \label{fig:subfig1vib}
  \end{subfigure}
  \caption{Comparison of the three procedures with respect to density plot of design points} 
    \label{fig:densityM3}
 \end{figure} 

Figure \ref{fig:densityM3} shows the density plot from all design points generated during the sequential stage for model $M_3$ using the three methods - PICS, balanced PICS and C-M for one single simulation. The true optimal design points are shown as vertical lines in both graphs. The initial design is uniform. It is seen that the design points generated by all three approaches converge to the true design as $n_1$ and $n$ increase. However, the PICS algorithms perform better than the C-M algorithm, especially for smaller $n$, and as expected, the balanced version of PICS allocates almost equal probabilities around the three optimal points.

Figure \ref{fig:efficiencyM2} and Figure \ref{fig:efficiencyM3} show the relative efficiency plots over time for the three methods for models $M_2$ and $M_3$ respectively. The plots for model $M_1$ are very similar to those for model $M_2$ and are therefore not displayed. All these plots show that the proposed PICS approach performs better than the C-M approach in terms of relative efficiency. As expected, at the end of the data generating process, all approaches achieve better efficiency under the simpler two-parameter model $M_2$ compared to the three-parameter model $M_3$. The balanced PICS approach works marginally better than the regular, unrestricted PICS approach. Whereas the initial design $\xi_0$ does not appear to impact the performance for model $M_2$, for model $M_3$, choice of a uniform design appears to have clear advantages over a 3-point design for the PICS approach, at least during the initial stages.

\begin{table}[ht]
\centering
 \begin{tabular}{| c | c | c | c| c | c | } 
    \hline
    Model & design & $\initialrun$ & $n$ & \multicolumn{2}{c|}{Median time(sec)}\\
    \cline{5-6} 
    & & & & CM & PICS  \\ 
    \hline
        \multirow{4}{*}{$M_1$} &  \multirow{2}{*}{Uniform} & 40 & 100 & 0.262 & 0.039  \\
            \cline{3-6}
            & & 60 & 200 & 0.999 & 0.12  \\
           \cline{2-6}
          &  \multirow{2}{*}{Three-point} & 40 & 100 & 0.265 & 0.04  \\
            \cline{3-6}
            & & 60 & 200 & 1.017 & 0.12  \\
           \hline  \hline
         \multirow{4}{*}{$M_2$} &  \multirow{2}{*}{Uniform} & 40 & 100 & 0.382 & 0.076  \\
            \cline{3-6}
            & & 60 & 200 & 1.461 & 0.222  \\
           \cline{2-6}
          &  \multirow{2}{*}{Three-point} & 40 & 100 & 0.397 & 0.08  \\
            \cline{3-6}
            & & 60 & 200 & 1.519 & 0.231 \\
           \hline \hline
          \multirow{4}{*}{$M_3$} &  \multirow{2}{*}{Uniform} & 40 & 100 & 1.708 & 0.142 \\
            \cline{3-6}
            & & 60 & 200 & 7.118 & 0.425  \\
           \cline{2-6}
          &  \multirow{2}{*}{Three-point} & 40 & 100 & 1.674 & 0.148  \\
            \cline{3-6}
            & & 60 & 200 & 7.111 & 0.431  \\
           \hline  
 \end{tabular}
  \caption{Comparison of the procedures on the basis of Median time}
 \label{table:comparisonnlm}
\end{table}

The resource saving ability of the PICS approach is also clear in these plots. For example, in Figure \ref{fig:efficiencyM2}(c), with a three-point initial design of size $n_1 = 60$, to attain an efficiency of 60\% for model $M_2$, the PICS algorithm (and its balanced version) need roughly $n=135$ trials, whereas the C-M algorithm requires approximately $n=185$ trials. With uniform initial designs (Figure \ref{fig:efficiencyM2}(d)), these numbers are lower -- approximately 120 for PICS and 160 for C-M.

\newpage

\subsection{Example 2: Logistic regression in factorial experiment} \label{ss:simul_glm}

Recall the $2^2$ factorial experiment described in Example \ref{example2} where a logistic model for the binary response $Y$ includes only the main effects of two input variables $X_1$ and $X_2$, as shown in (\ref{eq:logistic}). Here, the D-optimal design $\optdesign$ is obtained in the form of the proportions $\pcomp{r}{s} = \glmni{r}{s}/n$, where $n$ is the total number of experimental units and $\glmni{r}{s}$ denotes the number of units allocated to the $r$th level of $X_1$ and $s$th level of $X_2$, such that $\sum_{r,s} \glmni{r}{s} = n$, for $r,s = 1,2$. 

Under this setup, the total Fisher information matrix corresponding to $n$ units is calculated as $\xmatrix{n}^{\T} \weightmat{n} \xmatrix{n}$, where each row of the 
$n \times 3$ matrix $\xmatrix{n}$ is a three-dimensional vector $\left(1 , x_{1} , x_{2} \right)$ with $x_{1}$ and $x_{2}$ taking values $-1$ or $+1$ depending on the level combination that row represents. The matrix $\weightmat{n}$ is an $n \times n$ diagonal matrix depending on the unknown parameter $\bm{\theta} = (\beta_0, \beta_1, \beta_2)^\T$ of interest, with the diagonal entries calculated corresponding to each row of $\xmatrix{n}$ as $\pi(x_1, x_2) \left(1 - \pi(x_1, x_2) \right)$. The diagonal entries of $\weightmat{n}$ can take up any one of the four values $\wcomp{r}{s}$, for $r,s = 1,2$, corresponding to each of the four combinations of $x_1$ and $x_2$. 

\cite{Mandal2012} derived analytic closed form solutions for $\optdesign$ under special cases, conditioned on the values of $\vcomp{r}{s}$'s, where $\vcomp{r}{s} = 1/\wcomp{r}{s}$. These optimal solutions, presented as Corollary 1 and Corollary 2 in \cite{Mandal2012}, are stated below:

\begin{enumerate}[label=(\arabic*)]
    \item \label{D-opt Corollary 1} Corollary 1: Suppose $\vcomp{1}{2} = \vcomp{2}{1} = \vcomp{2}{2} = v$ and $\vcomp{1}{1} < 3v$, then the optimal solution is given by:
    \begin{equation}
     \poptcomp{1}{1} = \frac{3v - \vcomp{1}{1}}{9v - \vcomp{1}{1}} ,  \poptcomp{1}{2} = \poptcomp{2}{1} = \poptcomp{2}{2} = \frac{2v}{9v - \vcomp{1}{1}}. \label{eq:Mandal1}
     \end{equation}
    \item \label{D-opt Corollary 2} Corollary 2: Suppose $\vcomp{1}{1} = \vcomp{1}{2} = u$, $\vcomp{2}{1} = \vcomp{2}{2} = v$ and $u > v$, then the optimal solution is given by:
    \begin{equation}
     \poptcomp{1}{1} = \poptcomp{1}{2} = \frac{2u - v - d}{6(u - v)} ,  \poptcomp{2}{1} = \poptcomp{2}{2} = \frac{u - 2v + d}{6(u - v)}, \label{eq:Mandal2}
     \end{equation}
    with $d = \sqrt{u^2 - uv + v^2}$.
\end{enumerate}

The restrictions on $\vcomp{r}{s}$'s are satisfied by imposing equivalent restrictions on $\trueparm$. For instance, the conditions of the D-optimal design $\optdesign$ given in~\ref{D-opt Corollary 1} is satisfied if and only if 
\begin{equation}\label{glm_constrnts_1}
    \beta_0 = \beta_1 = \beta_2 \quad \text{and} \quad \lvert \beta_0 \rvert < c_0,
\end{equation}
where $c_0$ is approximately equal to $0.8314$. On the other hand, the conditions of $\optdesign$ given in~\ref{D-opt Corollary 2} is satisfied if and only if
\begin{equation}\label{glm_constrnts_2}
    \beta_2 = 0 \quad \text{and} \quad \beta_0 \beta_1 > 0.
\end{equation}
Keeping this in mind, we considered the following values of $\beta_i$'s in our simulation study:
\begin{enumerate}[label = (\alph*)]
    \item $\modeltrueparmval{\beta}{0} = \modeltrueparmval{\beta}{1} = \modeltrueparmval{\beta}{2} = 0.7125$, while using the closed form solution given in (\ref{eq:Mandal1}).
    \item $\modeltrueparmval{\beta}{0} = 1.5, \modeltrueparmval{\beta}{1} = 0.5, \modeltrueparmval{\beta}{2} = 0$, while using the closed form solution given in (\ref{eq:Mandal2}).
\end{enumerate}

We first implement the C-M method for the given setup using algorithm~\ref{fig:algCM}. Similar to the steps followed in Section~\ref{ss:simul_nlr}, we generate an initial set of designs of size $\initialrun$ and the corresponding responses, then calculate the MLE of $\trueparm$ by solving a constrained optimization problem, where the objective function is the likelihood function and constraints for cases~\ref{D-opt Corollary 1} and~\ref{D-opt Corollary 2} are respectively given in equations~(\ref{glm_constrnts_1}) and~(\ref{glm_constrnts_2}). We used the minimize function in python and applied the method of Sequential Least Squares Quadratic Programming (SLSQP) to solve the constrained optimization problem. Next, in the $i$-th step of the sequential stage, we obtain the design point $\mathbf{X}_i$ by maximizing a criterion function similar to (\ref{criterion}), given by
\begin{equation}
    \det \left [ \xmatrix{i-1}^{\T} \weightmat{i-1} \xmatrix{i-1} + w_i\cdot \mathbf{X}_i \mathbf{X}_i^{\T}  \right ], 
\end{equation}
where $\weightmat{i-1}$ is calculated based on the parameter estimate $\mle{i-1}$ and the previously collected data matrix $\xmatrix{i-1} \in \mathbf{R}^{(i-1) \times 3}$, and $w_i$ depends on $\mle{i-1}$ and $\mathbf{X}_i$. Consequently, we generate a new response and update the MLE of $\trueparm$.

Next, we implement the PICS method using algorithm~\ref{fig:algPICS}. First, we generate the initial set of design points in a similar way as in C-M. In the sequential stage, we obtain the new design point $\mathbf{X}_i$ in the $i$-th step by drawing an observation from the probability distribution associated with the locally D-optimal design $\optdesign$ at $\mle{i-1}$ given by (\ref{eq:Mandal1}) and (\ref{eq:Mandal2}), using probability-proportional-to-size (PPS) sampling. 


We use the same criteria outlined in Section~\ref{ss:simul_nlr} to compare the C-M and PICS approach in this section, given by:

\begin{enumerate}
\item The relative efficiency at step $i = 1, \ldots, n$, defined as
\begin{equation*}
    e_i = 1 - \left | \frac{ \det (\frac{1}{i} \xmatrix{i}^{\T} \weightmat{i} \xmatrix{i}) - \det (\mathbf{I}^*) }{\det (\mathbf{I}^*)} \right |,
\end{equation*}
 where $\weightmat{i}$ is calculated based on $\mle{i}$ and $\xmatrix{i}$, and $\mathbf{I}^*$ given by (\ref{eq:trueFI}) is the true Fisher information matrix at the true value of the parameter. We simplify $\mathbf{I}^*$ to get a much simpler expression of $\mathbf{I}^*$ as ${\mathcal{X}^*}^{\T} \mathbf{W}^* {\mathcal{X}^*}$, where $\mathcal{X}^*$ and $\mathbf{W}^*$ are given by
 \begin{equation*}
 \mathcal{X}^*= \begin{bmatrix}
                1 & 1 & 1 \\
                1 & 1 & -1 \\
                1 & -1 & 1 \\
                1 & -1 & -1
                \end{bmatrix}, \quad 
    \mathbf{W}^* = \text{diag}(\woptcomp{1}{1}\poptcomp{1}{1}, \woptcomp{1}{2}\poptcomp{1}{2}, \woptcomp{2}{1}\poptcomp{2}{1}, \woptcomp{2}{2}\poptcomp{2}{2})
\end{equation*}
Note that, $\woptcomp{r}{s}$'s and $\poptcomp{r}{s}$'s are calculated at the true value of the parameter.
Higher the $e_i$, closer is the estimated Fisher information till stage $i$ to the true Fisher information. 
\item The computing time required to generate $n$ design points.   
\end{enumerate}

\begin{figure}[!ht]
\centering
  \begin{subfigure}{0.5\textwidth}
    \centering
    \includegraphics[width=\linewidth]{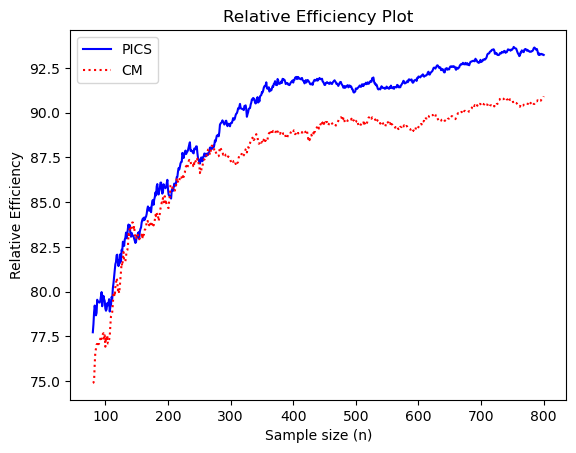}
    \caption{$n_1 = 80, n = 800$}
    \label{fig:subfig2ia}
  \end{subfigure}%
  \begin{subfigure}{0.5\textwidth}
    \centering
    \includegraphics[width=\linewidth]{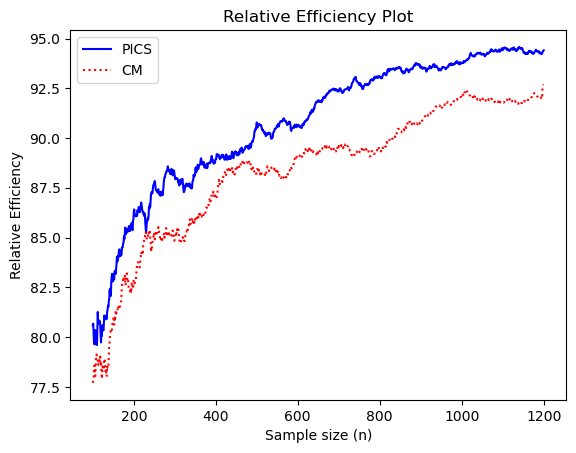}
    \caption{$n_1 = 100, n = 1200$}
    \label{fig:subfig2ib}
  \end{subfigure}
     \caption{Comparison of the three procedures with respect to efficiency for Case $1$} 
    \label{fig:efficiencyCor1}
\end{figure}

\begin{figure}[!ht]
\centering
  \begin{subfigure}{0.5\textwidth}
    \centering
    \includegraphics[width=\linewidth]{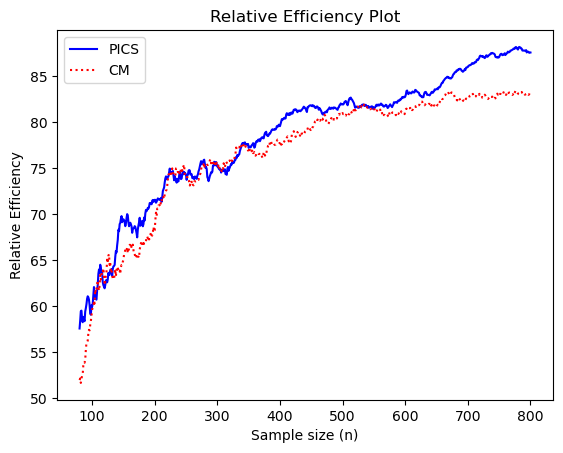}
    \caption{$n_1 = 80, n = 800$}
    \label{fig:subfig2iia}
  \end{subfigure}%
  \begin{subfigure}{0.5\textwidth}
    \centering
    \includegraphics[width=\linewidth]{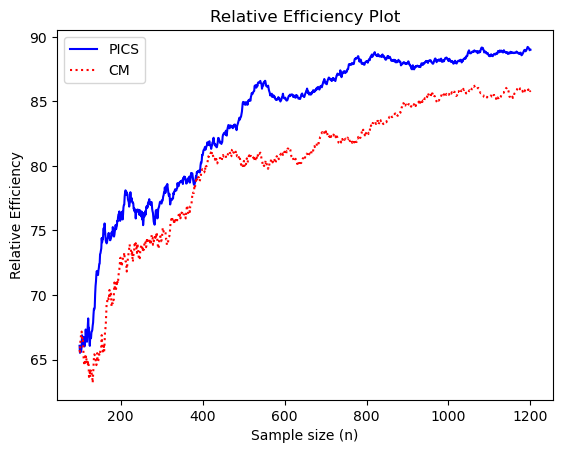}
    \caption{$n_1 = 100, n = 1200$}
    \label{fig:subfig2iib}
  \end{subfigure}
     \caption{Comparison of the three procedures with respect to efficiency for Case $2$} 
    \label{fig:efficiencyCor2}
\end{figure}

As in the previous section, the following two options of the initial and final design size $(n_1, n)$ have been considered:  (i) ($n_1 = 80, n=800$) and (ii) ($n_1 = 100, n=1200$). We consider the initial design $\xi_0$ as a four-point design, where each of the $2^2$ combinations of the experiment, i.e. each of the rows of $\mathcal{X}^*$ is drawn $n_1/4$ times each at any order. 50 simulations were conducted for each model under each setting. The efficiency curves in Figure \ref{fig:efficiencyCor1} and Figure \ref{fig:efficiencyCor2} and the computing times in Table \ref{table:comparisonglm} are based on means of these 50 simulations.

The relative efficiency plots over time are shown in Figures \ref{fig:efficiencyCor1} and \ref{fig:efficiencyCor2} for both C-M and PICS, corresponding to cases~\ref{D-opt Corollary 1} and~\ref{D-opt Corollary 2}, respectively. Overall, the plots suggest that the PICS method shows superior performance in terms of relative efficiency compared to the C-M method under the GLM setup as well. Table \ref{table:comparisonglm} shows that unlike the previous example, in this example the PICS approach is only marginally superior to the C-M approach in terms of computing time. The reason behind the two methods performing similarly in terms of computing time is that in this example, selection of new design points in the C-M approach essentially requires a search over a discrete space consisting of four candidate points.


\begin{table}[!ht]
\centering
 \begin{tabular}{| c | c | c | c |} 
    \hline
    Model & $\initialrun$ & \multicolumn{2}{c|}{Median time(sec)} \\
    \cline{3-4} 
    & & CM & PICS \\ 
    \hline 
      \multirow{2}{*}{Equation (\ref{eq:Mandal1})} & $80$  & $1.997$ & $1.592$\\
      \cline{2-4}
      & $100$  & $5.267$ & $3.904$  \\
    \hline \hline
      \multirow{2}{*}{Equation (\ref{eq:Mandal2})} & $80$  & $2.625$ & $2.423$ \\
      \cline{2-4}
      & $100$  & $5.922$ & $4.516$  \\
    \hline    
 \end{tabular}
 \caption{Comparison of the procedures on the basis of Median time}
 \label{table:comparisonglm}
\end{table}

\section{Concluding remarks and discussion} \label{sec:discussion}

In this paper, we propose an approach to enhance the efficiency of the standard sequential procedure for obtaining locally D-optimal designs for nonlinear models by leveraging existing closed-form solutions. The key idea is to ``plug-in'' the current parameter estimates into the available solutions, or alternatively to draw new design points from the known optimal design measure after substituting the current parameter estimates. We establish theoretically that the proposed procedure preserves the nice convergence properties of the standard approach and demonstrate its superiority with respect to efficiency and computation time via simulations. In addition to the two examples discussed in the paper, there are several other situations where closed-form solutions of D-optimal designs are available, making the proposed approach applicable to such settings. See, for example, \cite{Kalish1978}, \cite{Abdelbasit1983}, \cite{Haynes2018}.

A general problem associated with locally optimal designs is their reliance on assumed models. For instance, in Example \ref{example1} discussed in the paper, the experimenter would not know if the model $M_2$ or model $M_3$ explains the data generating mechanism better. One way to tackle such situations is to use model robust designs \citep[e.g.,][]{Dette1990, DetteFrank2001,Tsai2010, Smucker2011, Smucker2012, Smucker2015}. Adopting ideas from this literature on model robust designs to make the PICS approach more robust to model uncertainty would be an interesting extension of this work.

Bayesian sequential designs \citep{DrorStein2008} have been a popular alternative to the C-M type approach for non-linear models. This approach was used by \cite{LDHl2014} in the context of Example \ref{example1}. Integrating the PICS approach with the Bayesian framework would be an interesting possibility to consider and explore.

\newpage

\section{Appendix}
\label{sec:Appendix}

In this section we provide the detailed regularity conditions that are needed in our main Theorems~\ref{theorem1} and~\ref{theorem2}, as well as their proofs.

\subsection{Regularity conditions:}
\label{sec:regularity-conds}
\begin{enumerate}[label=(A.\arabic*)]
    \item~\label{condition:A.1} Given any $\bm{\theta}_1, \bm{\theta}_2 \in \bm{\Theta}$ define 
\begin{align}
    \EFish(\bm{\theta}_1, \bm{\theta}_2) = \Exs_{\mathbf{X} \sim \dist(\trueparm_2)} \left\{\Fish(\bm{\theta}_1, \mathbf{X})\right\} \label{eq:Jdef}
\end{align}
The function $\EFish(\trueparm, \cdot)$ is $L$-Lipschitz; i.e., for any $\bm{\theta} \in {\bm \Theta}$,
\begin{align}
\label{eqn:Lipschitz-property}
    \opnorm{\EFish(\bm{\theta}, \bm{\theta}_1) - \EFish(\bm{\theta}, \bm{\theta}_2)} &\leq L \|\bm{\theta}_1 - \bm{\theta}_2\|_2   \qquad \text{for all} \;\; \bm{\theta}_1, \bm{\theta}_2 \in \bm{\Theta} \subseteq \mathbf{R}^\dimn,  
\end{align}
where the subscript ``op'' represented operator norm.
   \item \label{Condition:A.2} (Control on the variance of the entries of $I(\bm{\theta}, \mathbf{X})$:) Let $[\mathbf{I}(\trueparm, \mathbf{X})]_{r,s}$ denote the $(r,s)^{th}$ entry of the matrix $\mathbf{I}(\trueparm,\mathbf{X})$, then we assume that variance of 
$[\mathbf{I}(\trueparm, \mathbf{X})]_{r,s}$ is uniformly bounded, i.e., 
\begin{align}
    \label{eqn:variance-control}
    \Var_{\mathbf{X} \sim \dist({\bm \theta}_1)} \left( [\Fish({\bm \theta}, \mathbf{X})]_{r, s} \right) \leq V_{\max} \qquad \text{for all} \;\; \trueparm \; \text{and} \; {\bm \theta}_1 \in {\bm \Theta} \subseteq \mathbf{R}^\dimn.  \;\; \text{and} \;\; 1 \leq r,s \leq \dimn. 
\end{align}

    \item \label{Condition:A.3} The function $\Fish(\cdot, \mathbf{x})$ is Lipschitz for every fixed $\mathbf{x} \in \exptspace$.
    
   \item \label{Condition:A.4} The sequence of estimators $\{\mle{n}\}$ generated through the PICS algorithm is a strongly consistent estimator of $\trueparm$.


    \item \label{condition:A.5} Support of $\likelihood$ does not depend on $\trueparm$ or $\mathbf{x}$. Also for fixed $\mathbf{x} \in \exptspace$ and $\outp \in \Rcal$, $\loglik$ is thrice continuously differentiable in $\trueparm$.
    \item \label{condition:A.6} Let $\score = \scorerhs$ be the gradient vector, that satisfies  
\begin{align*}
    \int_{\Rcal} \scorerhs \likelihood \mu (d\outp) = 0
\end{align*}
and 
\begin{align*}
    \sup _{\mathbf{x} \in \exptspace} \int_{\Rcal} \|\scorerhs\|_{\infty}^{2+t} \likelihood \mu (d\outp) < \infty
\end{align*} 
for some $t > 0$.
    \item \label{condition:A.7} Let $\hess$ denote the Hessian matrix (of order $\dimn$ x $\dimn$) of $\loglik$ corresponding to the second-order differentiation with respect to $\trueparm$. Then $\hess$ satisfies 
    \begin{eqnarray*}
        \int_{\Rcal} \hess \likelihood \mu (d\outp) 
        = & - \int_{\Rcal} \scorerhs \{ \scorerhs \}^{\T} \likelihood \mu (d\outp) = - \mathbf{I}(\bm{\theta}, \mathbf{x}),
    \end{eqnarray*}
    and 
    \begin{align*}
        \sup _{\mathbf{x} \in \exptspace} \int_{\Rcal}\|\hess\|_\infty^{2} \likelihood \mu (d\outp) < \infty
    \end{align*}
\item \label{condition:A.8} The Hessian matrix $H(\outp,\trueparm, \mathbf{x})$ is L-Lipschitz in $\trueparm$ for every fixed $\mathbf{x},y$.
    %
%

\end{enumerate}

\subsection{Proof of Theorem~\ref{theorem1}}
\label{sec:Proof-of-thm-1}

Throughout, we use $\mathbf{X}({\bm \theta})$ to denote a random sample drawn from  the distribution $\dist^*(\trueparm)$. Using definition (\ref{eq:Jdef}), $\EFish(\bm{\theta}, \bm{\thetahat}_{i - 1}) = \Exs_{\mathbf{X} \sim \dist^*(\bm{\thetahat}_{i - 1})} \left[ \Fish({\bm \theta}, \mathbf{X}) \mid \mathcal{F}_{i - 1} \right]$, where $\mathcal{F}_{i - 1}$ is the sigma-field generated by $(\points{Y}{i - 1}, \mathbf{X}_1, \ldots, \mathbf{X}_{i - 1})$, we have: 
\begin{align*}
    \frac{1}{n}\sum_{i = 1}^n \Fish(\bm{\theta}, \mathbf{X}_i(\bm{\thetahat}_{i - 1}))
    &= \left\{ \frac{1}{n}\sum_{i = 1}^n \Fish(\bm{\theta}, \mathbf{X}_i(\bm{\thetahat}_{i - 1})) - \frac{1}{n}\sum_{i = 1}^n \EFish(\bm{\theta}, \bm{\thetahat}_{i - 1}) \right\} + \frac{1}{n}\sum_{i = 1}^n \EFish(\bm{\theta}, \bm{\thetahat}_{i - 1}) \\
    &= T_1 + T_2,
\end{align*}
denoting the first and second terms by $T_1$ and $T_2$ respectively.

By definition, $\istar = \EFish(\bm{\theta}, \bm{\theta})$ and we have 
\begin{align}
\label{eqn:T2-bound-1}
    \opnorm{T_2 - \istar } &\leq  
    \frac{1}{n}\sum_{i = 1}^n \opnorm{\EFish(\bm{\theta}, \bm{\thetahat}_{i - 1}) -  \EFish(\bm{\theta}, \bm{\theta})} \notag  
    \leq \frac{L}{n} \sum_{i = 1}^n  \| \bm{\thetahat}_{i - 1} - \bm{\theta}\|_2,
\end{align}
where the last step follows from condition~\ref{condition:A.1}.

Using condition~\ref{Condition:A.4}, i.e., strong consistency of $\bm{\thetahat}_{i - 1}$ and using the fact that the average (cesaro-mean) of a convergent series of real number converges to the same limit, we conclude that $\opnorm{T_2 - \istar} \rightarrow 0$ almost surely. It remains to control the term $T_1$.  

\subsubsection*{Control on $T_1$:}
Note that $T_1$ is an average of Martingale difference sequence (MDS) with respect to the sigma-field $\mathcal{F}_{i - 1}$, and we aim to show that $T_1 \inProb 0$. We have  
\begin{align*}
    T_1 = \frac{1}{n}\sum_{i = 1}^n \left\{\Fish({\bm \theta}, \mathbf{X}_i(\thetahat_{i - 1})) -  \EFish(\bm{\theta}, \bm{\thetahat}_{i - 1}) \right\}
    \equiv \sum_{i = 1}^n \frac{\mathbf{Z}_i}{n} 
\end{align*}
Fix any entry $(p,q) \in [\dimn] \times [\dimn]$, where $\dimn$ is the dimension of the parameter $\trueparm$, and denote the $(p,q)^\text{th}$ entry of the matrix $\mathbf{Z}_i$ by $\mathbf{Z}_{i,p,q}$. Clearly, $\{ \mathbf{Z}_{i,p,q} \}_{1 \leq i \leq n}$ is a MDS with respect to the filtration $\{\mathcal{F}_{i}\}_{i \geq 1}$, the $\sigma$-field generated by the data. Fix any $\const > 0$, we have 
\begin{eqnarray*}
    \mathbb{P} \left( \left|\sum_{i \leq n} \mathbf{Z}_{i, p, q} \right| > n\const \right) 
    \leq \frac{\Exs \left(\sum_{i \leq n}\mathbf{Z}_{i, p,q}\right)^2}{n^2\const^2} 
    \stackrel{(i)}{=} \frac{\sum_{i\leq n} \Exs  \mathbf{Z}_{i, p,q}^2}{n^2\const^2}  
    \stackrel{(ii)}{\leq} \frac{nV_{\max}}{n^2\const^2} \rightarrow 0
\end{eqnarray*}
In other words, $\sum_{i \leq n} \mathbf{Z}_{i, p, q}/n \inProb 0$.
Step $(ii)$ above uses condition~\ref{Condition:A.2}. Step $(i)$ above uses the following standard identity for second moment of sum of MDS
\begin{align*}
    \Exs\left(\sum_{i \leq n} \mathbf{Z}_{i, p, q}\right)^2 = \sum_{i \leq n} \Exs \mathbf{Z}_{i, p, q}^2 + 2 \sum_{i < j} \Exs\Exs[\mathbf{Z}_{i, p, q} \cdot \mathbf{Z}_{j, p, q} \mid \mathcal{F}_{i - 1} ] = \sum_{i \leq n} \Exs \mathbf{Z}_{i, p, q}^2 
\end{align*}
Putting together the pieces we conclude $T_1 \inProb 0$ and $\opnorm{T_2 - \istar} \rightarrow 0$ almost surely. This completes the proof of (\ref{eq:theorem1_1}).

\subsubsection*{Proof of \eqref{eq:theorem1_2}:}
This part follows from condition~\ref{Condition:A.3}, i.e., the assumed Lipschitz property of $\Fish(\cdot, \mathbf{x})$. We have 
\begin{align*}
    \Big\| \frac{1}{n}\sum_{i = 1}^n \Fish(\bm{\thetahat}_n, \inp_i(\thetahat_{i - 1})) -  \frac{1}{n}\sum_{i = 1}^n \Fish(\bm{\theta}, \inp_i(\thetahat_{i - 1})) \Big\|_{\mathrm{op}}
    &\leq \frac{1}{n}\sum_{i = 1}^n 
    \Big\|\Fish(\thetahat_n, \inp_i(\thetahat_{i - 1})) - 
    \Fish(\trueparm, \inp_i(\thetahat_{i - 1})) \Big\|_{\mathrm{op}} \\ 
    &\leq   L \| \thetahat_n - \trueparm\|_2 \inProb 0
\end{align*}
The last line above uses condition~\ref{Condition:A.4}. This completes the proof of Theorem \ref{theorem1}.

\subsection{Proof of Theorem~\ref{theorem2}.}
\label{sec:Proof-of-thm-2}

We now establish the proof of Theorem~\ref{theorem2}. Let $\sigmafield{\designsize}{i}$ be an increasing sequence of $\sigma$-fields generated by $(\points{Y}{i}, \pointsbf{X}{i+1}$), $i \le \designsize-1$. Then, the gradient of the log-likelihood, i.e., $\gradient{i}{\designsize}$ is a sum of square integrable martingale difference sequence (MDS) with respect to the sigma-field $\sigmafield{\designsize}{i}$ by condition~\ref{condition:A.6}. Similarly, the sequence $\hessplusfish{i}{\designsize}$, i.e., the sum of the Hessian matrix of the log-likelihood and the Fisher information matrix is also a square integrable MDS  with respect to the sigma-field $\sigmafield{\designsize}{i}$ by condition~\ref{condition:A.7}. 

To establish the CLT for $\widehat{\bm \theta}_n$, we will prove the following:
\begin{eqnarray}
n^{-\frac{1}{2}} \gradient{i}{\designsize} &\converged& \mathcal{N}_\dimn \left (0, \istar \right ), \label{eq:CLT1}\\
\frac{1}{\designsize}\hessian{i}{\designsize} &\convergep& - \istar, \label{eq:CLT2}   
\end{eqnarray}
where $\istar =  \displaystyle \int_{\exptspace} \mathbf{I}(\trueparm, \mathbf{x}) \optdesign ( d \mathbf{x} ) $ is the true optimal Fisher information.

\newcommand{\real}{\mathbf{R}}


\subsubsection*{Proof of~\eqref{eq:CLT1}:}
Invoking the Cramer-Wold's theorem it suffices to show that for any vector $a \in  \real^\dimn$ with $\|a\|_2 = 1$  we have 
\begin{align}
    n^{-\frac{1}{2}} \sum_{i = 1}^\designsize a^\T G (Y_i , \trueparm , \mathbf{X}_i) \converged \mathcal{N} \left (0, a^\T\istar a \right ) 
\end{align}
It suffices to show that 

\begin{eqnarray}
\frac{\sum_{i = 1}^n (a^\T G(Y_i, \trueparm, \inp_i))^2 }{n}  \convergep a^\T \istar a, \quad \text{(Variance Stability)} \label{c2} \\
\frac{1}{n} \sum_{i = 1}^n \Exs \left[  (a^\T G(Y_i, \trueparm, \inp_i))^2 \cdot \mathbf{1}_{|a^\T G(Y_i, \trueparm, \inp_i)| > \sqrt{n} \const} \bigg| \mathcal{F}_{n, i-1} \right]  \convergep  0  \quad  \forall \const > 0, \quad \text{(Lindeberg Condition)} \label{c3}
\end{eqnarray}
To prove (\ref{c2}), note that
\begin{align*}
\sum_{i = 1}^n \frac{\Exs\left[G(Y_i, \trueparm, \inp_i) \{G(Y_i, \trueparm, \inp_i)\}^\T \mid \mathcal{F}_{n, i - 1} \right]}{n} \overset{(i)}{=} \frac{ \sum_{i = 1}^n \Fish(\trueparm, \mathbf{X}_i)}{n}  \overset{(ii)}{\convergep} \istar \quad, 
\end{align*}
where (i) follows from condition~\ref{condition:A.7} and (ii) from Theorem \ref{theorem1}. Next, to prove (\ref{c3}), note that for any $t>0$,
\begin{align*}
        & \frac{1}{n} \sum_{i = 1}^n \Exs \left[
        (a^\T G(Y_i, \trueparm, \inp_i))^2 \cdot \mathbf{1}_{|a^\T G(Y_i, \trueparm, \inp_i)| > \sqrt{n} \cdot \const} \bigg| \mathcal{F}_{n, i-1} \right]
        \\
        \le & \frac{1}{n} \sum_{i=1}^{\designsize} 
        \mathbb{E} \Big[ 
        (a^\T G(Y_i, \trueparm, \inp_i))^2 
        \cdot \mathbf{1}_{|a^\T G(Y_i, \trueparm, \inp_i)| > \sqrt{n} \cdot \const}
        \cdot \frac{\|G(Y_i, \trueparm, \inp_i)\|_\infty^t}{| \sqrt{n}\const |^t} \bigg| \sigmafield{\designsize}{i-1}
        \Big]\\
        \le & \frac{1}{\const^t n^{1+t/2}} \sum_{i=1}^{\designsize} \mathbb{E}\left[\| G(Y_{i}, \trueparm,\mathbf{X}_{i}) \|_\infty^{2+t} \bigg| \sigmafield{\designsize}{i-1} \right]\\
        \stackrel{(i)}{\leq} &  \frac{n}{\const^t n^{1+t/2}} \rightarrow 0 \quad [\text{as} \;\; n \rightarrow \infty].  
\end{align*}
where step~(i) follows from Condition~\ref{condition:A.6}. This completes the proof of claim~\eqref{eq:CLT1}. 

\subsubsection*{Proof of~\eqref{eq:CLT2}:}
By condition~\ref{condition:A.7}, $\{H(Y_i, \trueparm, \inp_i) + \Fish(\trueparm, \inp_i)\}_{i \geq 1}$ is a Martingale difference sequence. Also, by condition \ref{condition:A.7}, the second moment of entries of $H(Y_i, \trueparm, \inp_i)$ are uniformly bounded. Thus invoking the weak-law of large numbers~\cite[Theorem 2.13]{Hall1980}, we conclude 
\begin{equation}\label{hessfish}
    \frac{1}{\designsize}\hessplusfish{i}{\designsize} \convergep 0.
\end{equation}

We are now ready to complete the proof of Theorem~\ref{theorem2}. We have 
\begin{align*}
        0 & = \gradientest{i}{\designsize}\\ 
        & = \gradient{i}{\designsize} +  \left\{\sum_{i = 1}^n H(Y_i, \tilde{\trueparm}, \inp_i) \right\} (\mle{\designsize}-\trueparm) 
\end{align*}
where $\tilde{\trueparm}$ lies on the line joining $\trueparm$ and $\widehat{\trueparm}_n$. Rearranging the last relation we have that for $a \in \real^\dimn$ with $\|a\| \neq 0$,
\begin{align}
        \frac{1}{\sqrt{n}}\sum_{i = 1}^n  G(Y_i, \trueparm, \inp_i)
        &= \left\{\frac{1}{n} \sum_{i = 1}^n  -  H(Y_i, \trueparm, \inp_i) \right\} \sqrt{n} \cdot (\mle{\designsize}-\trueparm) + \sqrt{n} R (\mle{\designsize}-\trueparm)  \notag \\
        & \stackrel{(i)} = \big(\istar +  o_p(1) +  R \big) \cdot \sqrt{n} \cdot (\mle{\designsize}-\trueparm)
        \label{eqn:decomposition}
\end{align}
where step $(i)$ follows from~\eqref{hessfish} and the matrix $R$ is defined as 
\begin{align}
    R = \frac{1}{n} \sum_{i = 1}^n   \left\{ H(Y_i, \trueparm, \inp_i) - H(Y_i, \tilde{\trueparm}, \inp_i)  \right\}
\end{align}
Using the Hessian Lipschitz property~\ref{condition:A.8}, we have 
\begin{align}
\label{eqn:residual}
    \|R\|_{\mathrm{op}} &\leq \frac{1}{n}\sum_{i = 1}^n \| H(Y_i, \tilde{\trueparm}, \inp_i) - H(Y_i, \trueparm, \inp_i) \|_{\mathrm{op}} \leq  L\cdot \|\widehat{\trueparm}_n - \trueparm \|_2 \rightarrow 0 \quad \text{almost surely.}
\end{align}
Putting together the equations~\eqref{eqn:residual},~\eqref{eq:CLT1}, and~\eqref{eqn:decomposition} we conclude
\begin{align*}
    \sqrt{n}\cdot (\mle{\designsize}-\trueparm) 
    &= \big(\istar +  o_p(1) +  R \big)^{-1} \cdot \left\{ \frac{1}{\sqrt{n}}\sum_{i = 1}^n  G(Y_i, \trueparm, \inp_i)  \right\}  \\
    & \stackrel{d}{\rightarrow} \mathcal{N}_\dimn \left( 0, \istarinv \right)
\end{align*}
This completes the proof of Theorem~\ref{theorem2}.

\bibliographystyle{apalike}
\bibliography{SD_ref}

\end{document}